\documentclass[prd,preprint]{revtex4}
\usepackage{graphicx}
\usepackage{amssymb}

\begin{document}

\title{Comments on the Evolution of Strongly Degenerate Neutrinos in the Early Universe}
\author{Kazuhide Ichikawa}
\author{Masahiro Kawasaki}
\affiliation{Research Center for the Early Universe, University 
  of Tokyo, Bunkyo-ku, Tokyo 113-0033, Japan}
\date{\today}

\begin{abstract}
We reconsider the evolution of strongly degenerate neutrinos in the early universe. Our chief concern is the validity of the entropy conservation after the neutrino annihilation process has frozen out (so that the establishment of chemical equilibrium is not trivial). We argue that the entropy indeed conserves because elastic scattering keeps the neutrino and antineutrino distribution functions in the equilibrium form and the sum of their chemical potential keeps zero even after the neutrino annihilation freeze-out. We also simulate the evolution of the degenerate neutrino spectrum to support the argument. We conclude that the change in the neutrino degeneracy parameter when the relativistic degrees of freedom in the universe decreases is calculated using the entropy conservation and the lepton number conservation without worrying about at what temperature the neutrino annihilation process freezes out.
\end{abstract}
\maketitle

\section{introduction}
The cosmology with the strongly degenerate neutrinos or, in other word, large lepton asymmetry is one of well-investigated themes. There are researches concerning how to generate it and what kind of cosmological consequences it produces at various cosmological epochs (see Ref.~\cite{Dolgov2002}). In this paper, we make comments on thermodynamic properties of such degenerate neutrinos in the early universe. Some of them have been already pointed out in Refs.~\cite{Dolgov2002}\cite{Ichikawa2003} and we make complementary arguments. Others provide corrections to probable misunderstandings found in Ref.~\cite{Dolgov2002}. 

First of all, we would like to compare Refs.~\cite{Dolgov2002}\cite{Ichikawa2003} and this paper. In Ref.~\cite{Dolgov2002}, it is argued that the evolution of the neutrino degeneracy parameter $\xi_{\nu}$ is given by imposing the neutrino number conservation after the neutrino annihilation freeze-out. Therefore the freeze-out temperature and the variation of the temperature are regarded to be necessary, where the latter is calculated by integrating the covariant energy conservation law because the entropy conservation is not considered to hold. As for the antineutrino degeneracy parameter $\xi_{\bar{\nu}}$, the relation $\xi_{\nu}+\xi_{\bar{\nu}}=0$ is considered to break down. In Ref.~\cite{Ichikawa2003}, contrary to Ref.~\cite{Dolgov2002}, it is argued that the temperature variation can be calculated using the entropy conservation and that the relation $\xi_{\nu}+\xi_{\bar{\nu}}=0$ holds after the freeze-out. There, the entropy conservation is regarded to hold approximately and the justification for its application is attributed to the exponentially suppressed antineutrino number density under the existence of the strong degeneracy. In this paper, we note that the entropy conservation is derived using the relation $\xi_{\nu}+\xi_{\bar{\nu}}=0$, which holds also after the freeze-out as described in Ref.~\cite{Ichikawa2003}. Then the entropy conservation is valid at any time. Together with the lepton number conservation, we can calculate how the degeneracy evolves without knowing at what temperature the freeze-out takes place. This is the point we have missed in the previous paper and want to stress in this paper. However, we note that the calculations in Ref.~\cite{Ichikawa2003} are not wrong because the neutrino number conservation used there is virtually same as the lepton number conservation when the degeneracy is large. Just we have done redundant calculations on freeze-out temperature.

In Sec.~\ref{sec:entropy}, we first describe how the variation in the neutrino degeneracy parameter is calculated with the lepton number conservation assuming the entropy conservation. Then we show the entropy to conserve as long as neutrinos are kinematically coupled to the rest of the cosmic plasma. In Sec.~\ref{sec:simulation}, we show the results obtained by numerically simulating the degenerate neutrino spectrum. In this way, we directly confirm they are in thermal equilibrium and $\xi_{\nu}+\xi_{\bar{\nu}}=0$ holds. In Sec.~\ref{sec:conclusion}, we summarize the discussion and give the conclusion.

\section{the neutrino degeneracy variation calculated by the entropy conservation} \label{sec:entropy}

The lepton asymmetry in the universe is of course measured by the lepton number density: $n_l = n_{\nu}-n_{\bar{\nu}}$. When neutrinos and antineutrinos are in thermal equilibrium, their number densities are specified by temperature $T$ together with chemical potential $\mu$ (or as is frequently used, with degeneracy parameter $\xi=\mu/T$) and annihilation process $\nu+\bar{\nu}\leftrightarrow e^- +e^+$ ensures $\xi_{\nu}=-\xi_{\bar{\nu}}$ so that $n_l=\xi\{(\xi/\pi)^2+1\}T^3/6$ (we assume for simplicity $\xi_{\nu}>0$ and denote it as $\xi$). 

Since the lepton number $N_l=a^3 n_l$ and the total entropy $S=a^3 s$ in the universe conserve (where $a$ is the scale factor and $s$ is the entropy density), it is useful for quantifying the lepton asymmetry to consider their ratio $\eta_l \equiv n_l/s$, which also conserves. Using energy density $\rho$, pressure $P$ and number density $n$, $s$ is calculated as $s = (\rho+ P -\mu n)/T$, and is often written as $s = (2\pi^2/45) g_S(\xi,T) T^3$ where $g_s$ denotes the relativistic degrees of freedom. Then $\eta_l$ is related to the degeneracy parameter as, $\eta_l \propto (\xi^3+\pi^2\xi)/\sum g_S(\xi,T)$. 

It is obvious from the last expression that $\xi$ takes different values as the total relativistic degree of freedom of the universe changes and how much it does can be calculated with the $\eta_l$ conservation (especially, $\xi$ stays constant while total $g_S$ does not change and this is the reason that $\xi$ is often used to quantify the degeneracy).

However, when the degeneracy is very large, there are occasions that the entropy conservation is not trivial as explained below (the lepton number conservation evidently holds because it is respected in every relevant elementary process). With neutrino degeneracy, there exists more neutrinos and less antineutrinos than without it. Since this makes harder for neutrinos to find partners of annihilation process, it freezes out (the process rate becomes less than the cosmic expansion rate) at higher temperature. When the degeneracy is so large that neutrino annihilation freezes out before the muon and antimuon annihilate, we can not in general expect the entropy to conserve during the muon annihilation because the process which ensures the chemical equilibrium to hold has frozen out.

The higher neutrino annihilation freeze-out temperature under the existence of degeneracy has been noticed in Ref.~\cite{Freese1983} and its effect on cosmology has been discussed there and also in Ref.~\cite{Kang1992}. However, as pointed out by Ref.~\cite{Dolgov2002}, they have regarded the neutrino annihilation freeze-out as the neutrino decoupling so they have concluded that neutrinos would not be heated while the muon annihilation. This is not true because although neutrinos come short of annihilation partners, antineutrinos, they have enough elastic scattering partners, for example electrons. The correct picture is that even the neutrino annihilation freezes out, they keep contact with the rest of the cosmic plasma and decoupling does not take place. In other words, after the neutrino annihilation freeze-out, the chemical equilibrium in not ensured to hold but the kinetic equilibrium is.

In this picture, what occurs during muon annihilation is that neutrinos preserve number and keep equilibrium distribution with the temperature following photon's which falls more slowly than $a^{-1}$. That is,
\begin{eqnarray}
N_{\nu}=a^3 n_{\nu} \propto a^3 T^3 \int \frac{y^2}{\exp(y-\xi)+1}= const.
\end{eqnarray}
 while $aT$ deviates from unity and increases. Therefore, as is easily seen, $\xi$ has to decrease. On calculating how much $aT$ increases and $\xi$ decreases, Ref.~\cite{Dolgov2002} has suggested the use of covariant energy conservation law because they consider the entropy conservation does not hold when there is degeneracy. Also, they have stated that the relation $\xi_{\nu}+\xi_{\bar{\nu}}=0$, which is true during chemical equilibrium, breaks down. In summary, from Ref.~\cite{Dolgov2002}'s point of view, the evolution of $\xi$ would be calculated as: 1) calculate the neutrino annihilation freeze-out temperature, 2)use the total entropy conservation before the freeze-out, and 3)use the neutrino number conservation and the covariant energy conservation law after the freeze-out.
 
In our recent paper, we expressed different opinions from theirs. In Ref.~\cite{Ichikawa2003}, we argued that the total entropy conserves and $\xi_{\nu}+\xi_{\bar{\nu}}=0$ holds even after the freeze-out. Our argument is as follows.

The variation of the entropy is determined by the second law of thermodynamics \cite{EarlyUniverse}:
\begin{eqnarray}
TdS=d(\rho V)+PdV-\mu dN.
\end{eqnarray}
The first two terms on the right-hand side vanishes according to the covariant energy conservation. For the last term, there are contributions from neutrinos and antineutrinos so that
\begin{eqnarray}
TdS=-\mu_{\nu}dN_{\nu}-\mu_{\bar{\nu}}dN_{\bar{\nu}}=-(\mu_{\nu}+\mu_{\bar{\nu}})dN_{\nu}, \label{eq:dS}
\end{eqnarray}
where for the second equality, we use the lepton number conservation $dN_l=d(N_{\nu}-N_{\bar{\nu}})=0$. Before the neutrino annihilation freezes out, the particles are in chemical equilibrium so $\mu_{\nu}+\mu_{\bar{\nu}}=0$ holds and the entropy conserves even if $dN_{\nu} \neq 0$ (during for example muon annihilation). 

After the neutrino annihilation process has frozen out, since the chemical equilibrium breaks in general, we can not apply $\mu_{\nu}+\mu_{\bar{\nu}}=0$ immediately so it is not trivial that the entropy conservation holds when $dN_{\nu} \neq 0$. However, what we call here "freeze-out" is for neutrinos and not for antineutrinos who have a lot of annihilation partners. As a result, the chemical potential of antineutrinos is expected to keep the value $\mu_{\bar{\nu}}=-\mu_{\nu}$ and the entropy conserves.
 
To make this argument clearer, we consider the Boltzmann equation for the annihilation of antineutrinos during the muon annihilation. For that purpose, it is important to notice that even after the neutrino annihilation freeze-out, the elastic scattering is sufficiently frequent so every particle species are in kinetic equilibrium with certain well-defined temperature and as the universe expands, the kinetic equilibrium is maintained as the temperature decreases. The relevant part of the Boltzmann equation is
\begin{eqnarray}
\frac{d n_{\bar{\nu}}}{dt} +3 H n_{\bar{\nu}}&=&\int \frac{d^3p_1}{2E_1(2\pi)^3} \frac{d^3p_2}{2E_2(2\pi)^3} \frac{d^3p_3}{2E_3(2\pi)^3} \frac{d^3p_4}{2E_4(2\pi)^3}(2\pi)^4 \delta^{(4)}(p_1+p_2-p_3-p_4)  \nonumber \\
 & & \times |{\cal M}|^2 \bigg\{  [1-f_{\bar{\nu}}(E_1)][1-f_{\nu}(E_2)]f_{e^-}(E_3)f_{e^+}(E_4)  \nonumber \\
& &\hspace{3cm}- f_{\bar{\nu}}(E_1)f_{\nu}(E_2)[1-f_{e^-}(E_3)][1-f_{e^+}(E_4)] \bigg\},
\label{eq:nubarboltzmann}
 \end{eqnarray}
where $|{\cal M}|^2$ is the (angular integrated) invariant amplitude squared. Suppose that at first the universe has temperature $T_1$ with $\mu_{\nu}+\mu_{\bar{\nu}}=0$. At that moment, the number density obeys the Boltzmann equation without the expansion term $dn_{\bar{\nu}}/dt=0$. This is consistent with the collision term, the right hand side of Eq.~(\ref{eq:nubarboltzmann}), which vanishes when the distribution functions take equilibrium form $f(E)=1/[\exp\{(E-\mu)/T_1\}+1]$ with $\mu_{e^{\pm}}=0$, $\mu_{\nu}+\mu_{\bar{\nu}}=0$ and energy conservation $E_1+E_2=E_3+E_4$. As the muons annihilate and the universe expands, the equilibrium with temperature $T_1$ breaks but by the frequent elastic scattering, it quickly settles down to next equilibrium with $T_2 (<T_1)$. This transition is driven by the Boltzmann equation like Eq.~(\ref{eq:nubarboltzmann}) with electron and positron temperature $T_2$ and neutrino and antineutrino temperature $T_1$. The difference takes place because the electromagnetic interaction is much stronger than the weak interaction and electrons are immediately heated by annihilating muons but neutrinos are not. This gives non-zero collision term to evolve the distribution functions or, knowing they take equilibrium form, to evolve chemical potentials. The relation between the chemical potentials is determined by demanding the Boltzmann equation be $dn_{\bar{\nu}}/dt=0$ when $H\rightarrow 0$ and $T_1 \rightarrow T_2$. This is only achieved when the phase space factors on the right hand side of Eq.~(\ref{eq:nubarboltzmann}) cancel out i.e.~$\mu_{\nu}+\mu_{\bar{\nu}}=0$.

We would like to note three points. First, similar argument does not work with neutrino Boltzmann equation because after the neutrino annihilation freeze-out, when collision term is negligible to expansion term, it is $dn_{\nu}/dt + 3Hn_{\nu}\approx 0$ and does not give us information about distribution functions. Second, the argument above assumes the period in which temperature decrease is dictated by the muon annihilation in addition to the cosmic expansion. On the other hand, when particle degree of freedom is constant, the universe becomes cooler as a whole only by the cosmic expansion so there occurs no temperature difference between electrons and neutrinos. Then the terms in the right hand side of Eq.~(\ref{eq:nubarboltzmann}) cancel out to give the evolution equation $dn_{\bar{\nu}}/dt +3Hn_{\bar{\nu}}$ properly expressing the number conservation. Third, it is crucial that there exists temperature difference between electrons and degenerate neutrinos to show $\mu_{\nu}+\mu_{\bar{\nu}}=0$ but the difference is quickly erased due to the frequent elastic scattering and we can regard $T_{\nu}=T_{\gamma}$ for the cosmological time scale. More concretely, the argument in the previous paragraph is valid when $T_1-T_2$ is much smaller than the temperature difference between before and after the muon annihilation.

Now that we show $\mu_{\nu}+\mu_{\bar{\nu}}=0$, it is readily seen from Eq.~(\ref{eq:dS}) that the total entropy conserves even after the neutrino annihilation process freezes out. Then, the evolution of $\xi_{\nu}$ is calculated with the total entropy conservation and the lepton number conservation with $\xi_{\bar{\nu}}=-\xi_{\nu}$. Note that the neutrino annihilation freeze-out temperature is not necessary for the calculation, contrary to what is discussed in the literatures such as Refs.~\cite{Dolgov2002}\cite{Kang1992}\cite{Freese1983}. The naive treatment we have introduced at the beginning of this section turns out to be correct.

\section{numerical simulation of the neutrino spectrum evoltuion} \label{sec:simulation}
 
In this section, we simulate the evolution of degenerate neutrino spectrum to confirm the argument given in the previous section. We find manifestly the thermal equilibrium distribution is preserved with the same temperature as the photons and the relation $\xi_{\nu}+\xi_{\bar{\nu}}=0$ holds . 

Before showing the results, we describe our simulation method. Similar simulations are performed in Refs.~\cite{Dolgov1997}\cite{Esposito2000} and more details are found. We assume for simplicity that only electron-type neutrinos are degenerate and other types have no degeneracy. We evolve neutrino and antineutrino distribution functions, $f_{\nu}(y)$ and $f_{\bar{\nu}}(y)$, and the photon temperauture $T$ as functions of $x=m_0 a$ where $m_0$ is an arbitrary energy scale (we use $m_0=1$ GeV and set it unity hereafter). $y$ is defined by $y=a p$ where $p$ is the particle momentum. 

With these variables, derivatives of the distribution functions are calculated by
\begin{eqnarray}
\frac{df_{\nu(\bar{\nu})}}{dx}=\frac{C_{\nu(\bar{\nu})}}{Hx},
\end{eqnarray}
where $C$ is the collision term and $H$ is the cosmic expansion rate. $H$ is calculated from the total energy density $\rho_{tot}$ as $H=\sqrt{\rho_{tot}/3}/M_{pl}$ where $M_{pl}=2.436\times 10^{18}$ GeV is the Planck energy.

For the collision term, we include the elastic scattering $\nu(\bar{\nu})+e^{\pm} \leftrightarrow \nu(\bar{\nu})+e^{\pm}$ and the annihilation $\nu+\bar{\nu} \leftrightarrow e^- + e^+$. We denote the former contribution as $C^e$ and the latter $C^a$ so that $C=C^e+C^a$. We calculate them with approximation for electrons to be massless and to obey Boltzmann statistics as in Ref.~\cite{Kawasaki1999}. Then
\begin{eqnarray}
C^e_{\nu}(y)&=&\frac{2G_F^2[(C_V+1)^2+(C_A+1)^2]x}{\pi^3 y^2} \nonumber \\
& &\times \Bigg[ -f_{\nu}(y) \left\{ \int_0^y dy^{\prime} [1-f_{\nu}(y^{\prime})] F_1(y,y^{\prime}) +\int_y^{\infty} dy^{\prime} [1-f_{\nu}(y^{\prime})] F_2(y,y^{\prime}) \right\} \nonumber \\
& &\hspace{1cm}+[1-f_{\nu}(y)] \left\{ \int_0^y dy^{\prime} f_{\nu}(y^{\prime}) B_1(y,y^{\prime}) +\int_y^{\infty} dy^{\prime} f_{\nu}(y^{\prime}) B_2(y,y^{\prime}) \right\} \Bigg], \label{eq:cela}
\end{eqnarray}
where $G_F$ is the Fermi coupling constant, $C_V=-1/2$ and $C_A=-1/2+2\sin^2\theta$ ($\theta$: weak mixing angle). Functions $F$ and $B$ are defined by
\begin{eqnarray}
F_1(y,y^{\prime})&=&D(y,y^{\prime})+E(y,y^{\prime}) \exp\left(-\frac{y^{\prime}}{xT} \right), \\
F_2(y,y^{\prime})&=&D(y^{\prime},y) \exp \left( \frac{y-y^{\prime}}{xT} \right) + E(y,y^{\prime}) \left( -\frac{y^{\prime}}{xT} \right), \\
B_1(y,y^{\prime})&=&F_2(y^{\prime},y), \\
B_2(y,y^{\prime})&=&F_1(y^{\prime},y),
\end{eqnarray}
where 
\begin{eqnarray}
D(y,y^{\prime})&=&\frac{2T^4}{x^2} \left\{ y^2+y^{\prime 2} + 2(y-y^{\prime})xT +4 x^2 T^2  \right\}, \\
E(y,y^{\prime})&=&-\frac{T^2}{x^4} \left\{ y^2y^{\prime 2} + 2y y^{\prime}(y+y^{\prime})xT +2(y+y^{\prime})^2 x^2 T^2 + 4(y+y^{\prime})x^3 T^3 + 8x^4T^4 \right\}.
\end{eqnarray}

As for $C^a$, the expressions in Ref~\cite{Kawasaki1999} have not included neutrino degeneracy so we need to modify the annihilation term to be
\begin{eqnarray}
C^a_{\nu}(y)&=&-\frac{4G_F^2[(C_V+1)^2+(C_A+1)^2]}{9\pi^3 x^5} \nonumber \\
& &\hspace{1cm} \times\int dy^{\prime} y y^{\prime 3} \left\{ f_{\nu}(y)f_{\bar{\nu}}(y^{\prime}) -[1-f_{\nu}(y)][1-f_{\bar{\nu}}(y^{\prime})]\exp \left( -\frac{y+y^{\prime}}{xT} \right) \right\}. \label{eq:cann}
\end{eqnarray}

For antineutrinos collision terms, we have to just exchange $f_{\nu}$ and $f_{\bar{\nu}}$ in Eqs~(\ref{eq:cela}) and (\ref{eq:cann}).

The derivative of the temperature is obtained from the covariant energy conservation $d\rho/dx=-3(\rho+P)/x$. Since our simulation includes photons, electrons (approximated to be massless), muons, two types of neutrinos with no degeneracy and one type with degeneracy, 
\begin{eqnarray}
\frac{dT}{dx}&=&-\frac{\frac{1}{x}\left\{ 9\times 4\rho_{\gamma}+3(\rho_{\mu}+P_{\mu}) \right\}+ \frac{1}{2\pi^2 x^4} \int dy y^3 \frac{df_{\nu}}{dx} + \frac{1}{2\pi^2 x^4} \int dy y^3 \frac{df_{\bar{\nu}}}{dx} }{9\times \frac{\partial \rho_{\gamma}}{\partial T} + \frac{\partial \rho_{\mu}}{\partial T} }.
\end{eqnarray}

As the initial condition, we take $x=10$ which corresponds to $T=100$ MeV and assume degenerate neutrinos have equilibrium distribution with $\xi_{\nu}=-\xi_{\bar{\nu}}=10$. Actually, for this condition, since the annihilation rate exceeds cosmic expansion rate (see Ref.~\cite{Ichikawa2003}), neutrinos are in chemical equilibrium. The muon mass is about 106 MeV so they are almost fully relativistic (relativistic degree of freedom is 3.11, while fully relativistic particle would have 3.5) at 100 MeV. We follow the evolution down to until about 10 MeV at which temperature the muons almost annihilate away.

We note here some technical detail concerning numerical calculation.  To discretize momentum, we take equally spaced 100 points in $0<y<20$. Time step is fixed to $\Delta x =10^{-4}$. Since the differential equation for $f's$ are stiff but not for $T$, $f$'s are first evolved with 2nd order semi-implicit method and then, using that results, $T$ is evolved with 2nd order Runge-Kutta method.

\begin{figure}
\includegraphics{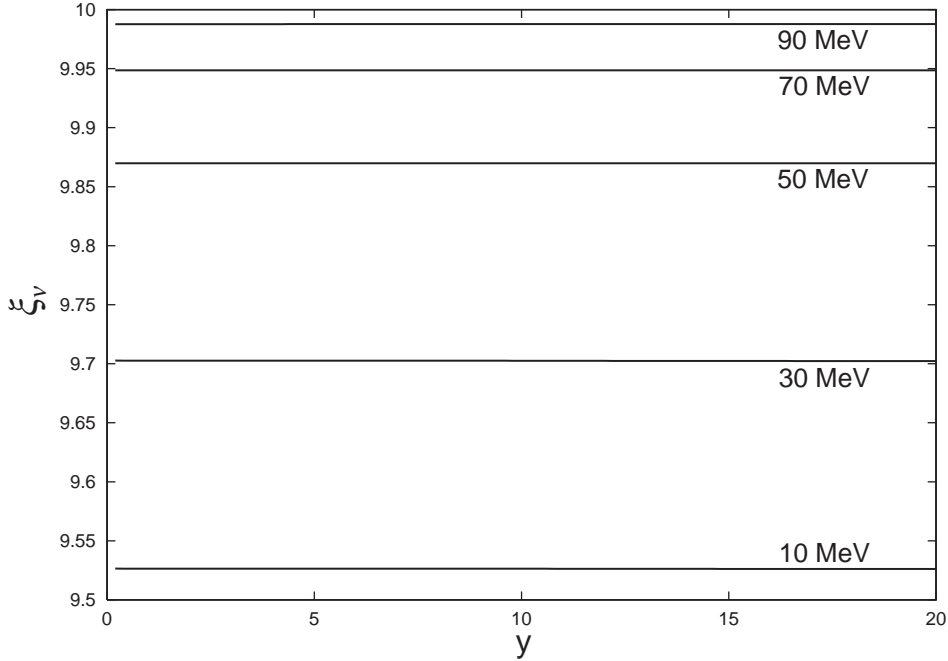}
\caption{This figure shows the evolution of $\xi_{\nu}(y)$. $\xi$ is computed as in the text. The appearance of the horizontal lines indicates that the neutrino distribution takes equilibrium form at each temperature. We plot $\xi_{\nu}$ at several temperatures. The greater line intervals correspond to the faster decrease in the muon relativistic degree of freedom.}
\label{fig:xi}
\end{figure}

\begin{figure}
\includegraphics{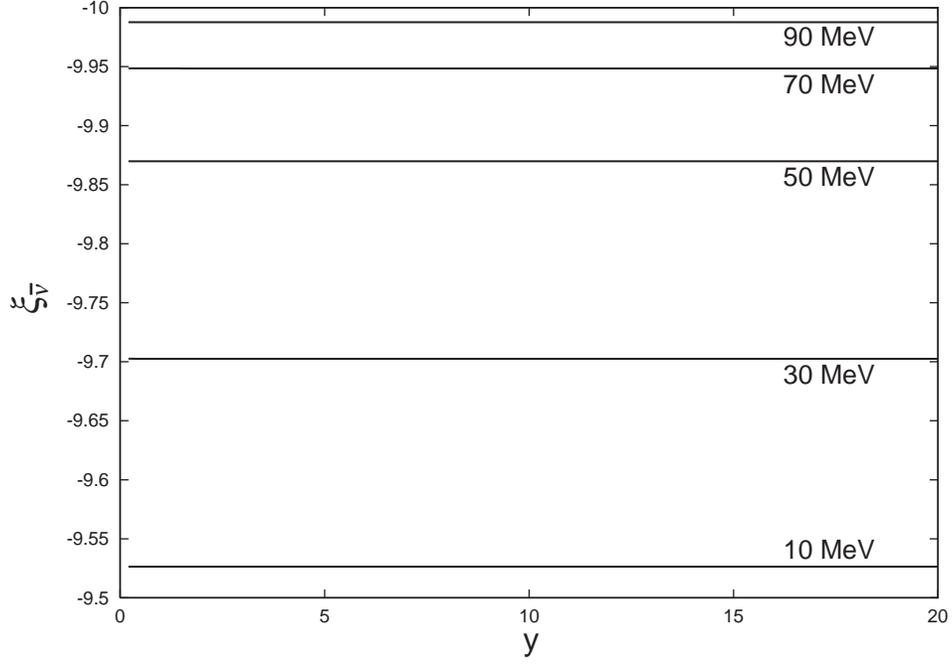}
 \caption{This figure shows the evolution of $\xi_{\bar{\nu}}$ with similar features to Fig~\ref{fig:xi}. We see $\bar{\nu}$ also keeps the equilibrium distribution. Together with Fig.~\ref{fig:xi}, they indicate the relation $\xi_{\nu} + \xi_{\bar{\nu}} = 0$ holds.}
\label{fig:xibar}
\end{figure}

\begin{figure}
\includegraphics{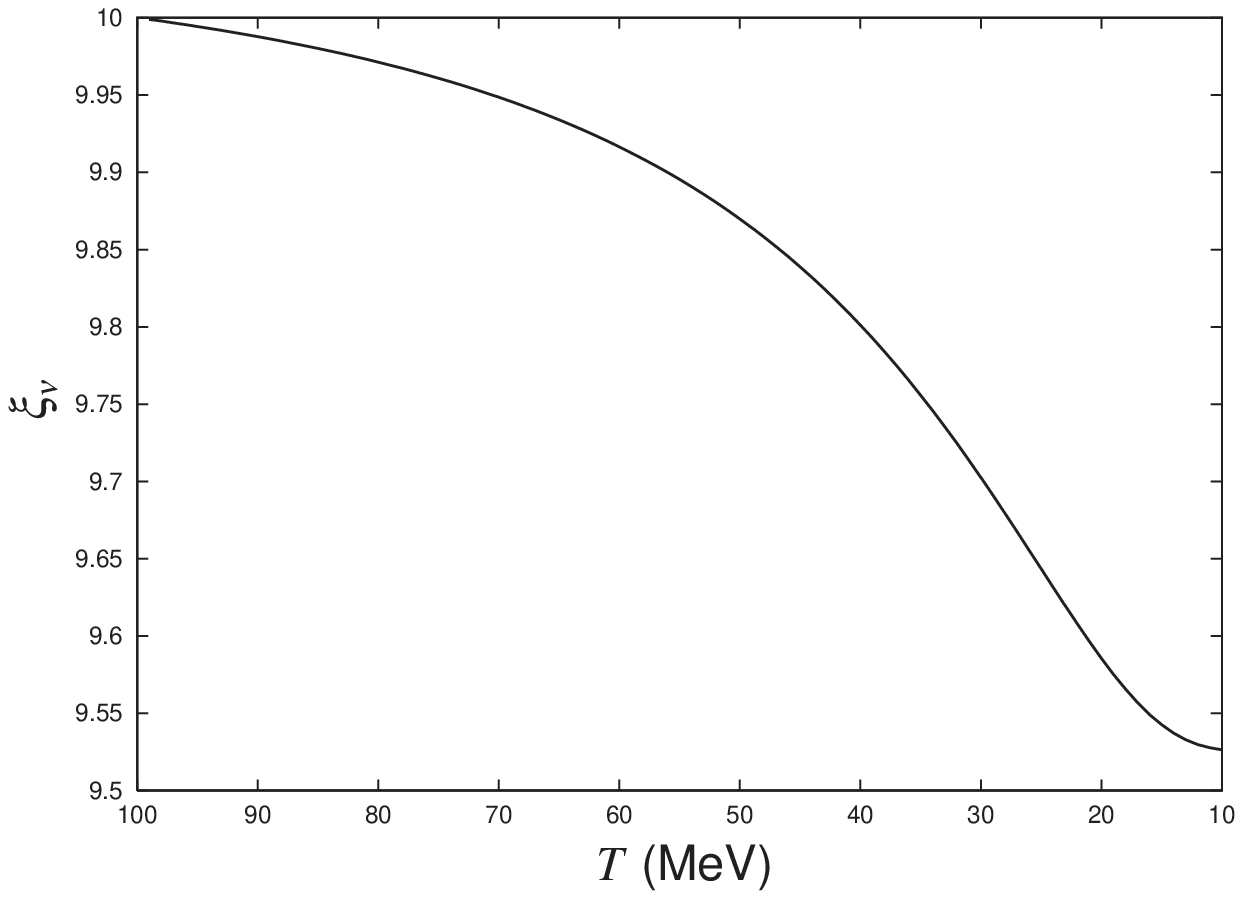}
\caption{The evolution of the neutrino degeneracy parameter. We plot $\xi_{\nu}$ at $y=7$ but it has practically no dependence on $y$ as shown in Fig.~\ref{fig:xi}. $\xi_{\bar{\nu}}$ evolves with opposite sign but the same absolute value.}
\label{fig:Txi}
\end{figure}

The simulation results are summarized in Figs.~\ref{fig:xi} to \ref{fig:Txi}. Figs.~\ref{fig:xi} and \ref{fig:xibar} indicate the spectra keep equilibrium form with well-defined temperature and chemical potential. Fig~\ref{fig:Txi} shows how the degeneracy parameter evolves which can be reproduced using the entropy conservation and the lepton number conservation.

For Figs.~\ref{fig:xi} and \ref{fig:xibar}, what we actually evolve is the distribution functions $f_{\nu}(y)$, $f_{\bar{\nu}}(y)$ and photon temperature $T$ but we express the results in terms of  momentum dependent degeneracy parameters $\xi_{\nu}(y)$ and $\xi_{\bar{\nu}}(y)$ which are calculated using the relations
\begin{equation}
f_{\nu(\bar{\nu})}=\frac{1}{\exp(\frac{p-\mu_{\nu(\bar{\nu})}}{T})+1}=\frac{1}{\exp(\frac{y}{xT}-\xi_{\nu(\bar{\nu})})+1},
\end{equation}
or $\xi_{\nu(\bar{\nu})}=y/(xT) -\ln(f_{\nu(\bar{\nu})}^{-1}-1)$. Note that we do not assume neutrinos and antineutrinos have equilibrium distribution at this stage. The result that $\xi$'s computed by such a way are independent of momentum $y$ (indicated by the horizontal lines which appear in these figures) tells the realization of the equilibrium distribution with the temperature $T$ and the degeneracy parameters $\xi_{\nu(\bar{\nu})}$.

In addition to $f_{\nu}$ and $f_{\bar{\nu}}$ are described by thermal equilibrium distribution with the same temperature as the photons, we see that the sum of their degeneracy parameters is accurately zero: $\xi_{\nu} + \xi_{\bar{\nu}}=0$ (to be more precise, the sum never exceeds $10^{-3}$). This relation is supposed to ensure the entropy conservation as discussed above so we should check whether the final value of the degeneracy parameter calculated by the numerical simulation is also obtained from the conservation laws. The relativistic degree of freedom of fermion with mass $m$ (ignoring spin and anti-particle) is
\begin{eqnarray}
g_{s,mass}=\frac{45}{\pi^2}\frac{s}{T^3}=\frac{45}{4\pi^4}\int_0^{\infty} dx \frac{x^2}{\exp (\epsilon)+1} \left( \epsilon + \frac{x^2}{3\epsilon} \right),
\end{eqnarray}
where $\epsilon=\sqrt{x^2+(m/T)^2}$. Then $g_{s,muon}$ decreases from $3.11$ at $T=100$ MeV to $0.00782$ at $T=10$ MeV. As for degenerate neutrinos, it is well approximated as $g_{s,\nu}=(7/4)\{(15/7)(\xi/\pi)^2+1\}$ and $g_{s,\bar{\nu}}=(45/2\pi^4)(4+\xi)e^{-\xi}$ so the latter can be neglected in the present case. The other particle species include photon, electron and two types of non-degenerate neutrinos ($g_{s,others}=9$). Requiring $\eta_l \equiv n_l/s \propto (\xi^3+\pi^2 \xi)/(g_{s,muon}+g_{s,\nu}(\xi)+g_{s,others})$ to conserve, $\xi$ is found to decrease from 10 to 9.526. This reproduces the numerical simulation results very well.

\section{Conclusion} \label{sec:conclusion}
In summary, under the existence of neutrino degeneracy in the early universe, we show the total entropy indeed conserves. This is not trivial after the neutrino annihilation process which establishes the chemical equilibrium freezes out. We argue that it does as long as the neutrinos are kinematically coupled to the rest of the plasma so that the relation $\xi_{\nu} + \xi_{\bar{\nu}} = 0$ holds. To confirm this argument, we simulate the degenerate neutrino spectrum evolution. As the result, we find neutrinos and antineutrinos have the thermal equilibrium distributions with the same temperature as the rest of the plasma and with the degeneracy parameters satisfying $\xi_{\nu} + \xi_{\bar{\nu}} = 0$. Consistently, we see the simulated evolution of the degeneracy parameters can be reproduced using the entropy conservation and the lepton number conservation. 

Neutrinos with so strong degeneracy that annihilation freezes out before muons annihilate have been thought to necessiate some special thermodynamic treatments since Ref.~\cite{Freese1983} and to these days. However, as we discussed in this paper, that is not necessary and the degeneracy parameter evolution is calculated using the entropy conservation and the lepton number conservation, no matter when their annihilation freezes out.

 \end{document}